\documentclass[12pt]{article}
\usepackage{graphicx}

\newcommand\pubnumber{}
\newcommand\pubdate{\today}

\def\iip{International Institute of Physics\\
Federal University of Rio Grande do Norte
Campus Universit\'ario, Lagoa Nova, 59078-970 Natal/RN, Brazil}

\def\Title#1{\begin{center} {\Large #1 } \end{center}}
\def\Author#1{\begin{center}{ \sc #1} \end{center}}
\def\Address#1{\begin{center}{ \it #1} \end{center}}

\newcommand\pubblock{\rightline{\begin{tabular}{l} \pubnumber\\
         \pubdate  \end{tabular}}}
\newenvironment{Abstract}{\begin{quotation}  }{\end{quotation}}
\newenvironment{Presented}{\begin{quotation} \begin{center} 
             PRESENTED AT\end{center}\bigskip 
      \begin{center}\begin{large}}{\end{large}\end{center} \end{quotation}}
\def\Acknowledgements{\bigskip  \bigskip \begin{center} \begin{large}
             \bf ACKNOWLEDGEMENTS \end{large}\end{center}}
\def\support{\footnote{RS acknowledges the support of FAPESP grant 2016/01343-7.}}

\def\beq{\begin{equation}}
\def\eeq#1{\label{#1}\end{equation}}
\def\eeqn{\end{equation}}

\def\beqa{\begin{eqnarray}}
\def\eeqa#1{\label{#1}\end{eqnarray}}
\def\eeqan{\end{eqnarray}}

\let\bar=\overbar

\def\Dslash{\not{\hbox{\kern-4pt $D$}}}
\def\dslash{\not{\hbox{\kern-2pt $\del$}}}

\def\msb{{\bar{\ssstyle M \kern -1pt S}}}

\usepackage{amssymb}

\newcommand{\pa}[1]{\left(#1\right)}

\begin{document}
\begin{titlepage}
\pubblock

\vfill
\Title{Gravitational Waves and Neutrinos}
\vfill
\Author{Riccardo Sturani\support}
\Address{\iip}
\vfill
\begin{Abstract}
We give an overview about the recent detection of gravitational waves by the
Advanced LIGO first and second observing runs and by Advanced Virgo, with emphasis on the
prospects for multi-messenger astronomy involving neutrino detections.
\end{Abstract}
\vfill
\begin{Presented}
NuPhys2017, Prospects in Neutrino Physics\\
Barbican Centre, London, UK,  December 20--22, 2017
\end{Presented}
\vfill
\end{titlepage}
\def\thefootnote{\fnsymbol{footnote}}
\setcounter{footnote}{0}

\section{Introduction}
The detection of GW170817 \cite{TheLIGOScientific:2017qsa}, the coincident
Gamma Ray Burst (GRB) \cite{Monitor:2017mdv}, and the other electro-magnetic
counterparts in a wide region of the spectrum from X to radio frequencies
\cite{GBM:2017lvd} marked the historical debut of Gravitational Waves (GWs)
on the stage of Multi-messenger Astronomy in the first month of joint activity of
the Advanced LIGO \cite{TheLIGOScientific:2014jea} and Advanced Virgo detector
\cite{TheVirgo:2014hva}.

Historically, the first extra-solar multi-messenger detection happened with SN1987A
\cite{Kunkel:1987zz} and involved a wide spectrum of electromagnetic signals and
neutrinos that were detected by Kamiokande II \cite{Hirata:1987hu}, the
Irvine-Michigan-Brookhaven Experiment(IBM) \cite{Bionta:1987qt} and Baksan
\cite{Alekseev:1988gp} neutrino observatories with energy $\sim 20$ MeV.

More recently, another multi-messenger transient event involving neutrinos has been detected,
namely the extremely-high-energy neutrino event EHE170922A ($\sim$ PeV) \cite{ehenu} by
IceCube and consistent detections of gamma ray flares by Fermi-LAT \cite{eheFermi} and MAGIC
\cite{eheMagic} in a window of $\pm$ 5 days enabled to identify the source as
the active galactic nucleus TXS 0506+056 of the blazar category
(i.e. with its relativistic jet directed to Earth
producing a characteristic radio emission spectrum),
with yet undetermined red-shift.
Current interpretation of this astrophysical event is of ultra-high-energy
protons producing pions eventually decaying into gamma rays and neutrinos.

In general powerful astrophysical objects, endowed with magnetic field
and in the presence of strong astrophysical shocks are capable to accelerate particles
to extremely high energy cosmic rays, whose subsequent interaction with 
radiation and matter can produce electromagnetic waves in an wide frequency range,
and high energy neutrinos through decay of mesons.

Neutrinos can reach Earth from the entire cosmos because of their weak
interaction with matter (and null interaction with electromagnetic radiation).
Even more so for GWs, which can also be originated by the coherent, accelerated
motion of astrophysically massive objects.
As neutrinos and GWs can travel almost unaltered (apart from cosmological red-shift)
from sources to detectors, they represent invaluable messengers to bring a snapshot of the source
at their production.

Active searches for joint detections of GW events in coincidence with electromagnetic
and neutrino events have received strong momentum by the first detection of GWs from
binary black holes GW150914 \cite{Abbott:2016blz}, see
\cite{Abbott:2016gcq} for a search of an electromagnetic counterpart and
\cite{Adrian-Martinez:2016xgn} for a neutrino one. 

As in this first GW observation, neither neutrinos nor electro-magnetic counterparts
have been detected in coincidence with the additional GW events produced by
binary black holes coalescences so far observed by LIGO:
GW151226 \cite{Abbott:2016nmj}, GW170104 \cite{Abbott:2017vtc},
GW170608 \cite{Abbott:2017gyy} and by LIGO and Virgo GW170814
\cite{Abbott:2017oio}:
the negative results of searches for coincident neutrino events are
reported in \cite{ANTARES:2017iky}, \cite{Albert:2017obm}, \cite{Aab:2016ras}.

Note that since the decay of a binary black hole pair is very slow, it is
not expected to be surrounded by matter at the time of coalescence, thus
preventing the non gravitational messengers to be triggered by these kind of events,
although exotic mechanisms to produce multi-messenger signals from binary black
hole coalescences have been conceived, see e.g. \cite{Perna:2016jqh}, \cite{Bartos:2016dgn}.

Combined searches for transient events by GW and neutrino detectors have been
realized in the
pre-GW-detection era as well, see e.g. \cite{AdrianMartinez:2012tf} for an early
attempt and \cite{Aartsen:2014mfp} for a search of GWs in coincidence with
neutrino events in a period of joint operations between year 2007 and 2010.
Note that the 37 high-energy neutrino events at energies $20$ TeV $-$ PeV of cosmic origin
observed by IceCube in a 3 year period 2010 and 2013 \cite{Aartsen:2014gkd} have no astrophysical
association. In particular 2 of the $E> 100 $ TeV neutrinos
were detected during the nominal initial LIGO-Virgo observation periods, however not
all of the three GW detectors
were operational, thus jeopardizing the possibility to determine interesting
limits on joint sources of GWs and high energy neutrinos.

Upper limit rates of combined GW and high energy neutrino ($\gg$ GeV) source population on Initial (Advanced) LIGO-Virgo real (projected) data were computed in
\cite{Bartos:2011aa}, with the result of non-detections implying
less than $10^0-$few$\times 10^{-2}$ event per Milky Way equivalent galaxy per year 
for isotropic equivalent energy release in the range $10^{-2} -10^{-4}$ $M_\odot$ 
(and upper limits on the event number almost 2 order of magnitude stricter for Advanced LIGO-Virgo).

The outline of this short paper is the following: in sec. \ref{sec:Sources} we give
an overview of standard, non-exotic sources for a joint GW-neutrino detection and in
sec.~\ref{sec:gwdet} we summarize the salient features of the sources that are
imprinted in a GW observations and the prospect for future detections.
We conclude in sec.~\ref{sec:Conclusions} with an outlook of future combined detection of GWs and neutrinos.

\section{Potential sources for coincident neutrino and gravitational detections}
\label{sec:Sources}

Main candidates for a future detection of GWs and neutrinos are GRBs from core
collapse supernova and compact coalescing binaries involving at least one non-black hole object.

The duration distribution of GRBs shows a clear bi-modality, enabling the
distinction between short ($\lesssim 2$ sec) and long ones ($\gtrsim 2$ sec),
both being distributed isotropically in the sky supporting a cosmological origin
and the former being preferentially harder than the latter, see
e.g.~\cite{Berger:2013jza}.

Short GRBs were widely believed to originate from coalescing binaries involving
at least one neutron star already before GW170817 detection, model that this
observation has confirmed. Neutrino emission models cannot exceed (per flavor)
fluence value
\cite{ANTARES:2017bia}
\begin{equation}
F\simeq \rm{few}\times \pa{\frac E{\rm{GeV}}}^{-2}\pa{\frac d{40\,\rm{Mpc}}}^{-2}
\rm{GeV}^{-1}\rm{cm}^{-2}\,,
\end{equation}
for the most optimistic choice of parameters \cite{Kimura:2017kan}
(and on-axis view, for a $\pm 500$ sec window around trigger time), with maximal neutrino energy expected between $100$ TeV and $10$ PeV,
and neutrinos can be emitted within minutes of the GRB since they are associated
with both prompt and extended emission of gamma rays. Even more
stringent fluence upper limit applies if a millisecond, rapidly spinning, highly magnetized neutron star
(\emph{magnetar}) is created by the binary neutron
star merger, which can emit over a longer time span (weeks)
\cite{Fang:2017tla} (and for which maximal neutrino energy can exceed $100$ PeV).
In the case of maximal fluence, this limit is scraping the bottom of the upper limit
set by the ANTARES, IceCube and Auger non-observation of a neutrino counterpart
of GW170817 \cite{ANTARES:2017bia}, which were able constrain the 
neutrino fluence in the $100$ GeV $-$ $10^5$ PeV energy window, with GW170817 having
origin in a galaxy at a distance of $40$ Mpc from Earth.

Neutrino flux can also be modulated by the viewing angle, with fluxes
decreasing rapidly when the observation angle exceeds the opening angle of the
jet.

The energy released in GWS by short GRB events originated by coalescing binaries can extend up to $10^{-2}\,M_\odot$, from which GW spectral amplitude $h_{s}$ can be computed\footnote{For a signal with time domain profile $h(t)$ one can define the
root-sum-square amplitude $h_s\equiv\int_{-\infty}^\infty dt h_+^2(t)+h_\times^2(t)$.}
\begin{equation}
h_s\simeq 10^{-21} {\rm Hz}^{-1/2} \pa{\frac E{10^{-2}M_\odot}}^{1/2}
\pa{\frac{f_{GW}}{1\,{\rm kHz}}}^{-1}\pa{\frac D{1\,\rm{Mpc}}}^{-1}\,,
\end{equation}
allowing a detection up to $O(100)$ Mpc given the noise level of Advanced interferometer
that have reached sensitivity $h_s\sim$ few $\times 10^{-23}$ Hz$^{-1/2}$ in the frequency range between $100$ Hz and $1$ kHz.

Pre-GW detection era results tried to dig into GW data to search for signals in
coincidence with both and long GRBs \cite{Abbott:2016cjt}, with the
result that no association could be made, with closest GRB considered 
being at a distance
$\sim 150$ Mpc (corresponding to red-shift $z\simeq 0.05$).

Long GRBs are instead thought to be originated by the core-collapse of massive
(few $\times 10\,M_\odot$) stars \cite{MacFadyen:1998vz} and their isotropically 
equivalent luminosities can range between $10^{51}$ and $10^{53}$ erg/sec.
According to the collapsar model long GRBs are triggered by the core-collapse
explosion of a stripped-envelope massive star, 
after which powerful jets of matter plow through the collapsing star along the
spin-axis, eventually matter flows towards a newly formed black hole or magnetar 
reaching relativistic speeds, and producing GRBs.

High energy neutrinos ($\gg$ GeV) can be produced before the jet outflows and
their production can still be active during the afterglow, so that no clear
prediction can be made for their arrival time.
Neutrino fluxes for energies in the range $100$
GeV to $100$ TeV can be estimated to be around
$100\times (d/10\, \rm{Mpc})^{-2}$ events at a distance $d$
(for a km-scale water- or ice-Cherenkov detector) \cite{Hummer:2011ms}, where
value at least $O(10)$ are phenomenologically interesting.

Long GRBs can are further divided on observational basis into low-luminosity
core collapse supernovae, engine-driven core collapse supernovae, standard
low-luminosity GRBs with core-collapse supernovae, and canonical long GRBs,
with considerable uncertainty in estimates for neutrino fluxes.

Another interesting sources for joint detection of GWs and neutrinos are
magnetars which represent the best model to explain soft gamma repeaters and
anomalous X-ray pulsars, which display repeated outburst of short duration
($\sim 0.1 $ sec) with peak luminosity of $\sim 10^{42}$ erg/sec, thus much
less luminouos than GRBs (even though some rare SGR have
reached luminosities of $10^{47}$ erg/sec \cite{Hurley,Mereghetti}).
LIGO/Virgo have already searched in 2008-2010 data for signals in coincidence
with galactic soft gamma repeaters, with negative results \cite{Abadie:2010wx}.

\section{Gravitational Wave detector physics}
\label{sec:gwdet}

Advanced LIGO and advanced Virgo GW detectors are Michelson interferometer with Fabri-Perot cavities which represent
the most precise ruler ever made: by measuring the differential variation of
the interferometer's arms they can monitor the passage of a GWs in the frequency
range from few tens of Hz to roughly $1$ kHz. Because of the frequency range
interferometric GW detectors
are sensitive only to binary coalescence of \emph{compact} objects, thus
small enough ($\sim 10-100$ km) that can achieve such high orbital frequencies.
Interferometers respond linearly to the GW strain by measuring the difference in
optical path with the result of being mild directional detectors, as they can
detect only GWs that do not alter simmetrically the two end mirrors.

\begin{figure}[htb]
\centering
\includegraphics[width=.4\linewidth]{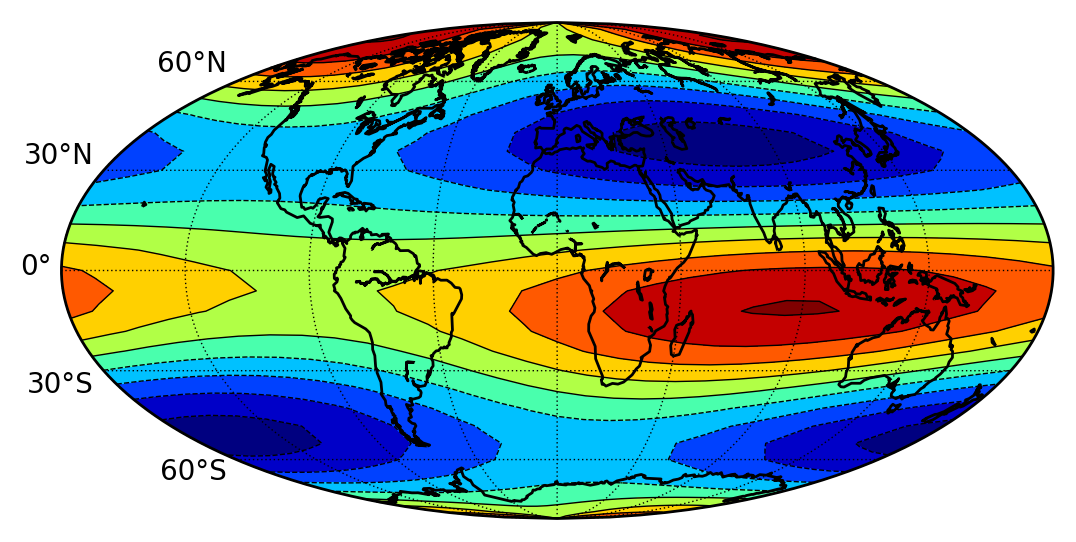}
\includegraphics[width=.4\linewidth]{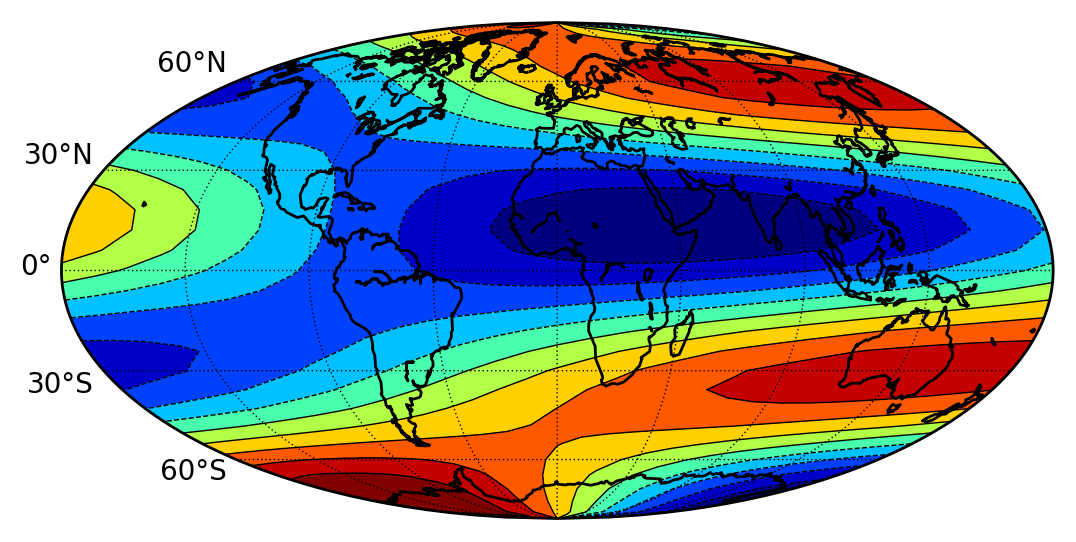}\\
\includegraphics[width=.4\linewidth]{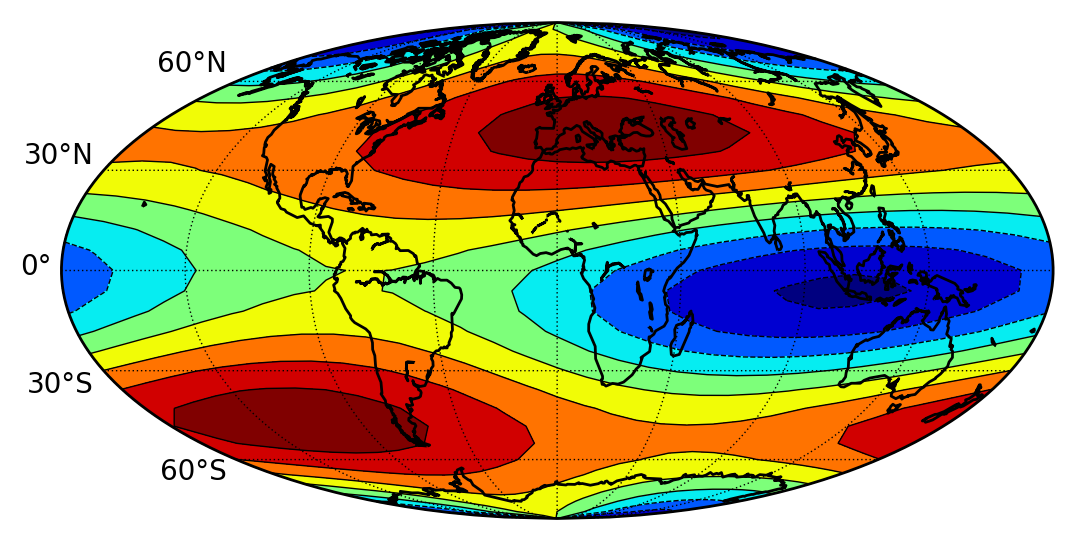}
\includegraphics[width=.4\linewidth]{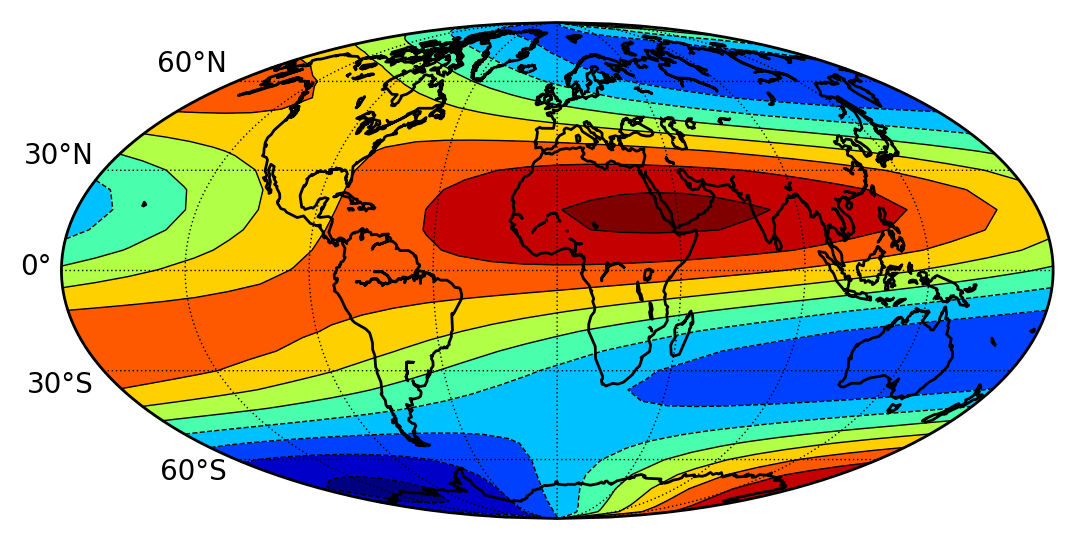}\\
\includegraphics[width=.4\linewidth]{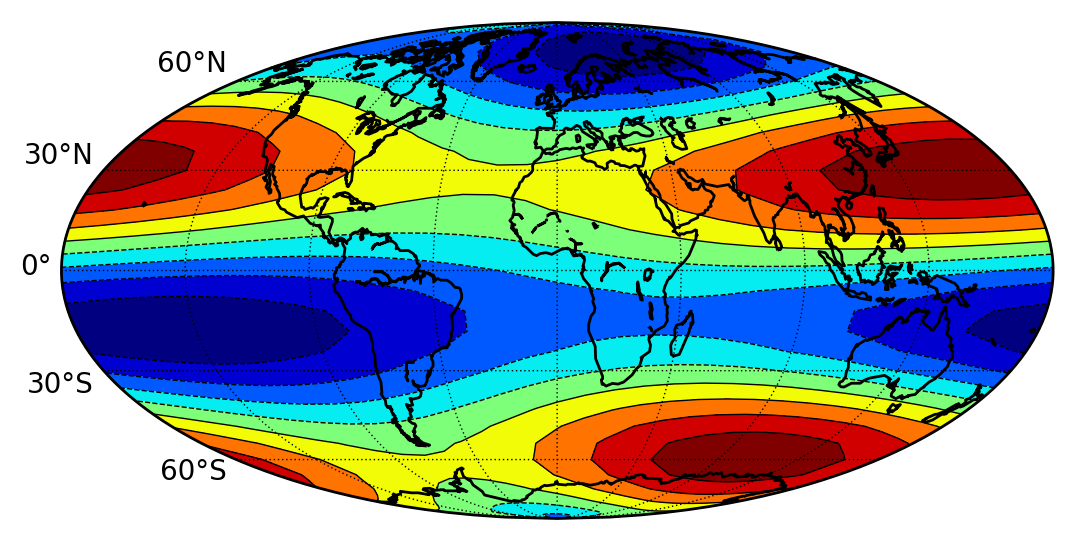}
\includegraphics[width=.4\linewidth]{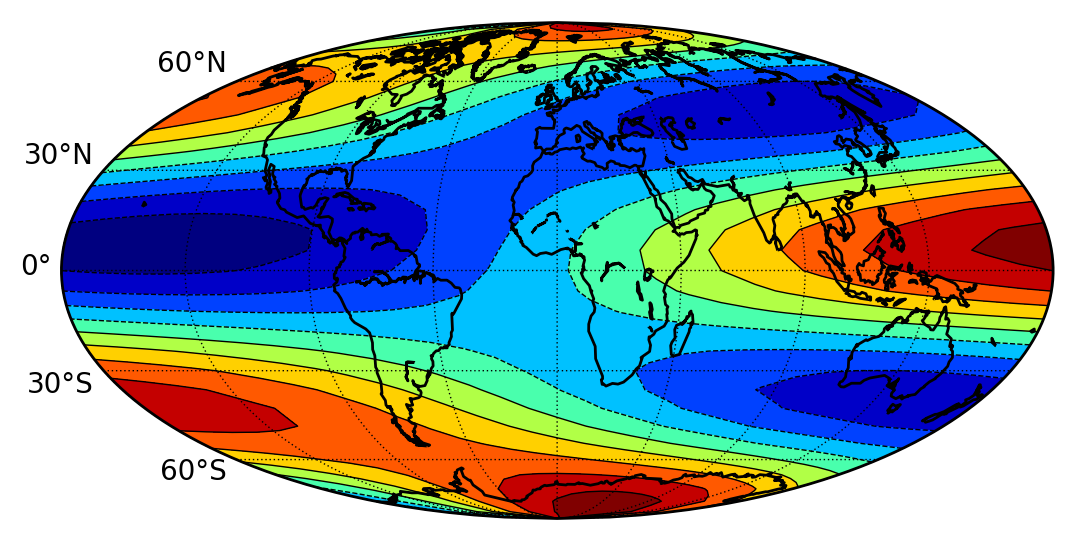}
\caption{Pattern functions of the LIGO Hanford (first line), LIGO Livingstone
(second line) and Virgo detector (third line) as a function of right ascension
and declination at the time of GW170817: August
17th 2017, 12:41:53 UTC. The first and second column represents respectively
$F_+$ and $F_\times$, the position of the GW170817 source being right ascension$=13h\,09'\,48''$, declination$=-23^o\,22'\,53''$. Pattern function values range
from 1 (dark red) to -1 (dark blue).}
\label{fig:patterns}
\end{figure}

GWs have 2 polarizations, conventionally called $h_+$ and $h_\times$ and
each detector is sensitive to only one linear combination of them, the
coefficients of proportionalitys between detector output and $h_{+,\times}$ being
the \emph{pattern functions} $F_{+,\times}$, see fig.~\ref{fig:patterns} for the values
of the LIGO and Virgo pattern functions at the time of GW170817.

For un-modeled events LIGO and Virgo search for excess noise but for coalescing
binaries accurate theoretical models exist enabling to correlate observational
data with pre-computed templates.

One important quantitative detail is that because of the
quadrupolar nature of the source the two polarization are affected in a
specific way by the relative orientation of the binary orbital plane and the
observation direction. Denoting such angle by $\iota$ one has
\begin{equation}
\begin{array}{rcl}
h_+&\propto& (1+\cos^2\iota)/2\,,\\
h_\times&\propto& \cos\iota\,,
\end{array}
\end{equation}
introducing a degeneracy between $\iota$ and the source-observer distance
to which the GW amplitude is inversely proportional: stronger signals
could equally well be closer and misaligned or farther and better aligned, with 
the latter possibility favoured a priori because at a larger
distance more volume is available, hence more possible sources \cite{Schutz:2011tw}.

\begin{figure}[htb]
\centering
\includegraphics[width=.6\linewidth]{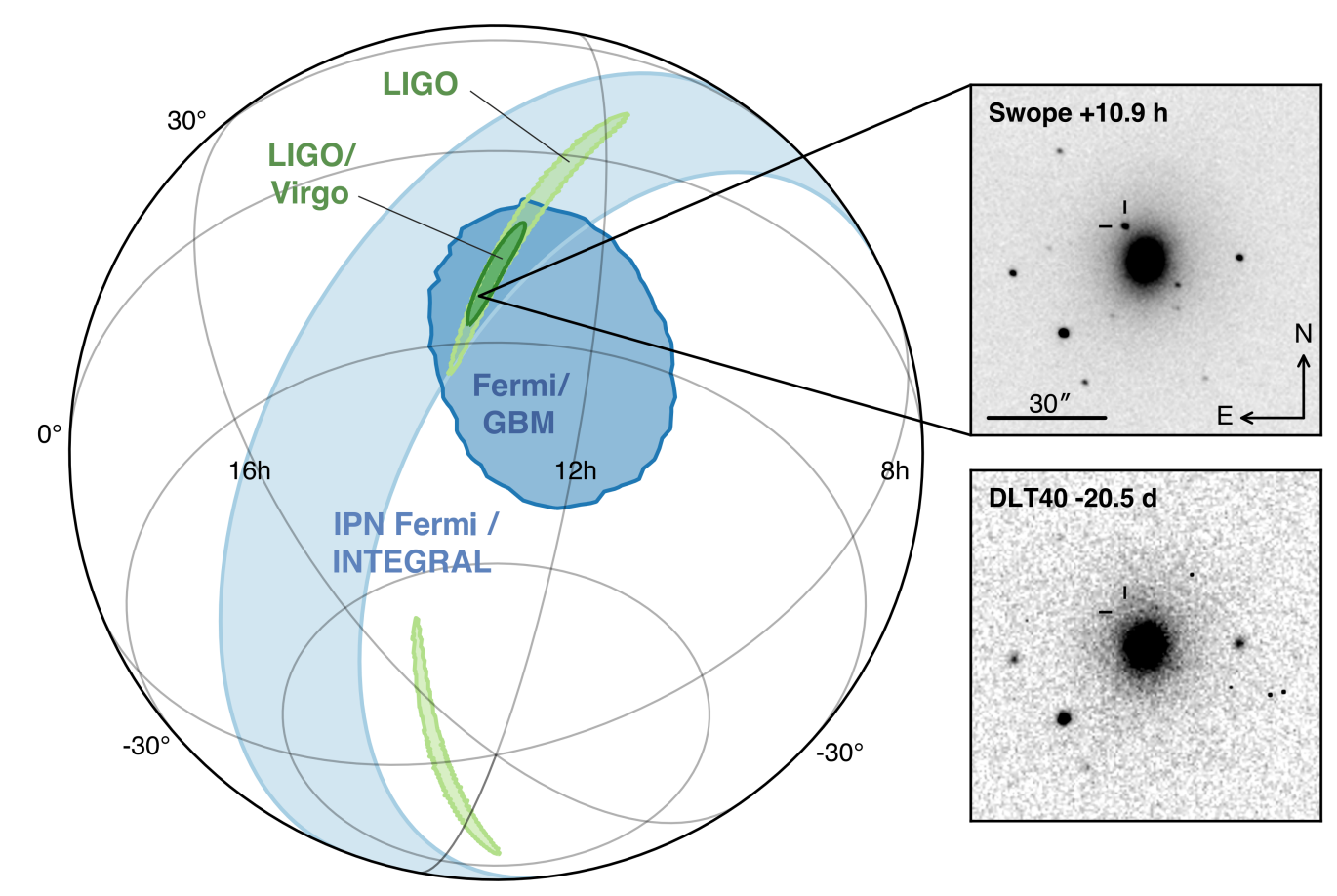}
\caption{LIGO/Virgo, Fermi and optical transient sky localization of respectively GW170817/GRB 17817A/SSS17a/AT 2017gfo. From \cite{GBM:2017lvd}.}
\label{fig:local}
\end{figure}

GWs can be localized with reasonable accuracy (e.g. the 90\% credible region
of GW170817 wich happened at 40 Mpc from Earth and was observed by 3
detectors, measured $28$ degree squared, with lower precision expected for
fainter objects) by short-circuiting the data of the time of arrival
(triangulation) and the information from the signal amplitudes and phases across
the detector network \cite{Aasi:2013wya}, with the result shown in
fig.\ref{fig:local} for GW170817, where the GRB \cite{GBM:2017lvd} and optical
\cite{Coulter:2017wya} localizations are also shown.

\begin{figure}
\centering
\includegraphics[width=.6\linewidth]{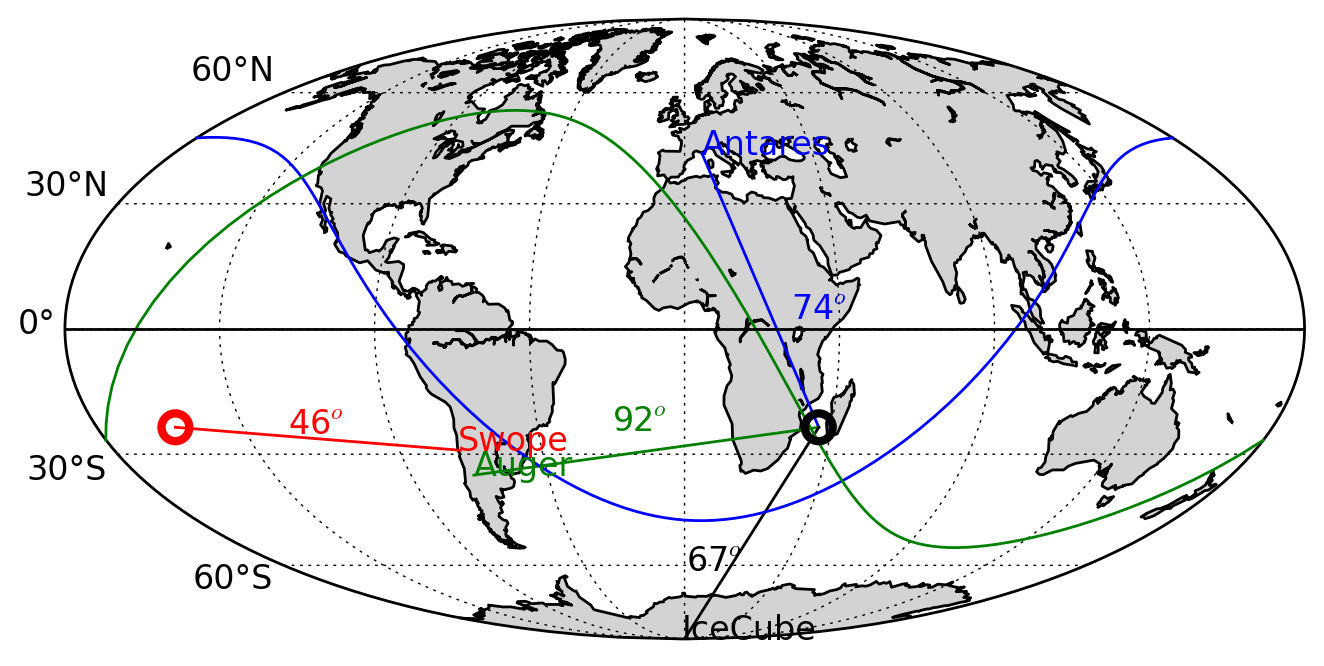}
\label{fig:nuorient}
\caption{Location of neutrino experiments and line dividing up/down directions.
Also reported are the position of GW170817 at emission (black circle) and at
the time of the first optical detection (red circle), 10.87h afterwards, by the One-Meter Two-Hemisphere
(1M2H) team with the 1 m Swope telescope at Las Campanas Observatory in Chile.
Adapted from \cite{ANTARES:2017bia}.}
\end{figure}

Source-detector orientation also matters for neutrino observatories
as for e.g. surface neutrino detectors like Auger have the optimal condition
represented by high-energy ($\gg 100$ TeV) showers created close to the
detector by neutrinos with earth-grazing incidence, see fig.~\ref{fig:nuorient}
for the neutrino detector orientation with respect to GW170817.
In-ice (like IceCube) and in-water (like ANTARES) detectors can take advantage of earth shield for up-going
neutrino (however Earth stop being transparent at $E>10$ PeV)
or look preferably at sufficiently high energy ($>100$ TeV) to kill the
background of penetrating muons from cosmic ray showers.

\section{Conclusions}
\label{sec:Conclusions}

The study of transient sources, which involve compact objects and ultra-violent 
phenomena (such as gamma-ray bursts and magnetars) is a very active,
promising and new field of reserch in astronomy.
Neutrino and gravitational waves have their own specific characteristics
making them unique messengers from the most energetic astrophysical events,
beside carrying invaluable information on the fundamental structure of matter
and interactions.

Detectors sensitivities are expected to improve in the coming years, pushing
for a perseverant effort in the coordination for a joint detection, possibly solving in
the future the puzzle of the still unexplained origin of the extremely high
energy neutrino events ($\sim$ PeV) \cite{Aartsen:2013bka}, which could be the
first indication of an astrophysical neutrino flux, and allowing another
insightful messenger to carry information about astrophysical sources.

\vspace{.5cm}

\Acknowledgements
RS thanks the organizer of the NuPhys2017 conference for the invitation and
helpful discussion that occurred in the meeting.

\end{document}